**Discovery methods for systematic analysis of causal molecular networks in modern *omics* datasets**

**Jack Kelly[1], Carlo Berzuini[1], Bernard Keavney[2,3], Maciej Tomaszewski[2,4], Hui Guo[1]**

[1] Centre for Biostatistics, School of Health Sciences, Faculty of Medicine, Biology and Health, University of Manchester, Manchester, UK, [2] Division of Cardiovascular Sciences, Faculty of Medicine, Biology and Health, University of Manchester, Manchester, UK, [3] Division of Cardiology and Manchester Academic Health Science Centre, Manchester University NHS Foundation Trust, Manchester, UK, [4] Manchester Heart Centre and Manchester Academic Health Science Centre, Manchester University NHS Foundation Trust, Manchester, UK

Corresponding author: Jack Kelly (jack.kelly@manchester.ac.uk)

**Funding**

This work was jointly supported by the British Heart Foundation and The Alan Turing Institute (which receives core funding under the EPSRC grant EP/N510129/1) as part of the Cardiovascular Data Science Awards (Round 2, SP/19/10/34813).

**Abstract**

With the increasing availability and size of multi-omics datasets, investigating the casual relationships between molecular phenotypes has become an important aspect of exploring underlying biology and genetics. This paper aims to introduce and review the available methods for building large-scale causal molecular networks that have been developed in the past decade. Existing methods have their own strengths and limitations so there is no one best approach, and it is instead down to the discretion of the researcher. This review also aims to discuss some of the current limitations to biological interpretation of these networks, and important factors to consider for future studies on molecular networks.

**Introduction**

Molecular networks are important to understanding biological process beyond the analysis of a single gene or molecule (1). The operation of molecular phenotypes at all levels is not isolated and interactions make up complicated networks that contain a wealth of information. In an age where data is being produced more than ever, these networks can become increasingly complex. A molecular network contains a set of nodes and edges. Nodes represent information from multi-omics, including but not limited to genes, messenger RNAs (mRNAs), proteins, DNA methylation patterns and protein phosphorylation. Edges represent the relationship between the nodes and so can symbolise direct and indirect relationships between molecular phenotypes and transcriptional regulation.

One of the primary advantages of molecular networks is in elucidating genetic and biological mechanisms underlying disease. Even in diseases with known causative genes (eg. CFTR mutation causing Cystic fibrosis (2) and mutations in *HTT* leading to Huntington's disease (3)) these genes act as part of a large network and never in isolation. Dysregulated biological processes and important 'hubs' within them can be identified as disease drivers, which potentially help identify drug targets that impact sets of associated genes rather than important individual genes, though this has yet to be translated to clinically useful therapies (4).

Many undirected networks (as shown in Figure 1A) rely on using correlation between nodes to infer symmetric associations. However, causal networks aim to differentiate the directed regulatory relationships from just associations. This approach identifies directed (as shown in Figure 1B) or mixed networks (as shown in Figure 1C). It is worth noting that directed relationships in a network do not necessarily have a causal interpretation, as they may merely depict temporal orders in the data generating process. Only if the confounders between the nodes have been adjusted for will these relationships have a causal meaning.

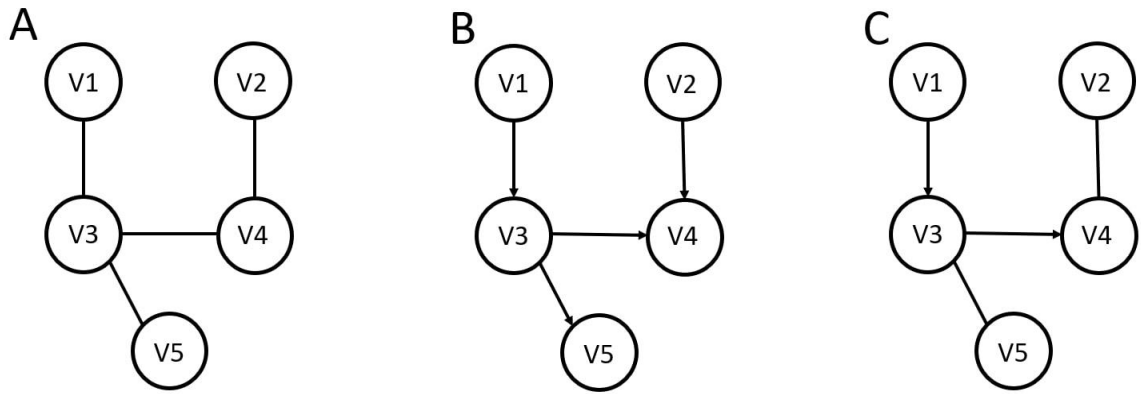

**Figure 1.** A) An example of an undirected network, B) a directed network and C) a mixed network. Mixed networks have both directed and undirected edges.

Identifying causal relations from gene expression data was proposed over 20 years ago (5) and since then a large number of causal inference methods have been developed using omics data. This approach is advantageous in the study of biology as it allows for inferring causality without interventions, especially when randomised controlled trials are infeasible due to high cost and ethical issues (6).

As the technology becomes more accessible, there is an increasing range of omics data that is being collected, which allows for integrative analysis to develop a more complete picture of how different types of omics interact with one another (7). Causal inference in molecular networks is a growing area of research. However, complex high dimensional causal networks have limited use and their contribution to the literature is heavily restricted as they are often difficult to interpret. There needs to be approaches that allow for identification of biologically important sub-networks and a small number of targets for future research or therapeutic intervention.

In this review, we will discuss the current literature using causal discovery methods on molecular networks and challenges that the area is facing. We will also discuss factors that influence interpretation of causal networks, including clustering and visualisation. Previous reviews (8,9) have focussed on introducing methodologies of building causal networks and given few biological examples, however here we will focus on published methods and their applications specifically to molecular networks and subsequent biological interpretation.

**Undirected networks**

Undirected networks have been an important approach for the investigation of biological processes and identification of hub genes in disease. Traditionally, protein-protein interaction networks have been built using a combination of *in vivo* and *in vitro* methods to understand interactions, however these approaches have huge time and financial costs, and result in noisy networks with many false positives (10). Approaches to omics data using *in silico* methods have been used as an alternative to better understand these undirected associations.

Most commonly, co-expression molecular networks are built on the basis of co-expression, usually measured using correlation (11). Transcriptomics data, using mRNAs to measure gene expression, is being increasingly generated. It has become popular to use specific R software to infer undirected networks from such data. For example, weighted gene co-expression network analysis (WGCNA) (12) is particularly user-friendly as the authors have produced extensive tutorials and guides to increase accessibility to researchers. Gene co-expression networks have successfully been used to identify important gene clusters (modules) and hub genes in many diseases, including cancer (13) and neurodegenerative disease (14). In many cases, the genes within these modules are investigated to see if any biological pathways are enriched using freely available tools such as the Gene Ontology (GO) (15,16) and Kyoto Encyclopedia of Genes and Genomes (KEGG) (17) databases.

Novel network approaches to hub detection (14) have been developed to understand important genes in disease. Although providing limited mechanistic understanding, undirected networks are important as they are often precursors of the study of causal networks.

**Causal molecular networks**

Applications of different causal methods to omics data is covered in this review. The simplest causal network only involves the causal relationship between a pair of variables, investigating whether a single exposure can cause a single outcome. Causal networks can be made increasingly complex to investigate the relationships between thousands of variables. Here, we consider Mendelian randomisation (MR), Bayesian networks (BN) and the PC algorithm that have been used individually with molecular phenotype data, as shown in Figure 2. We then focus on ensemble approaches to reduce the limitations of any single method. A summary of the methodologies discussed here are shown in table 1.

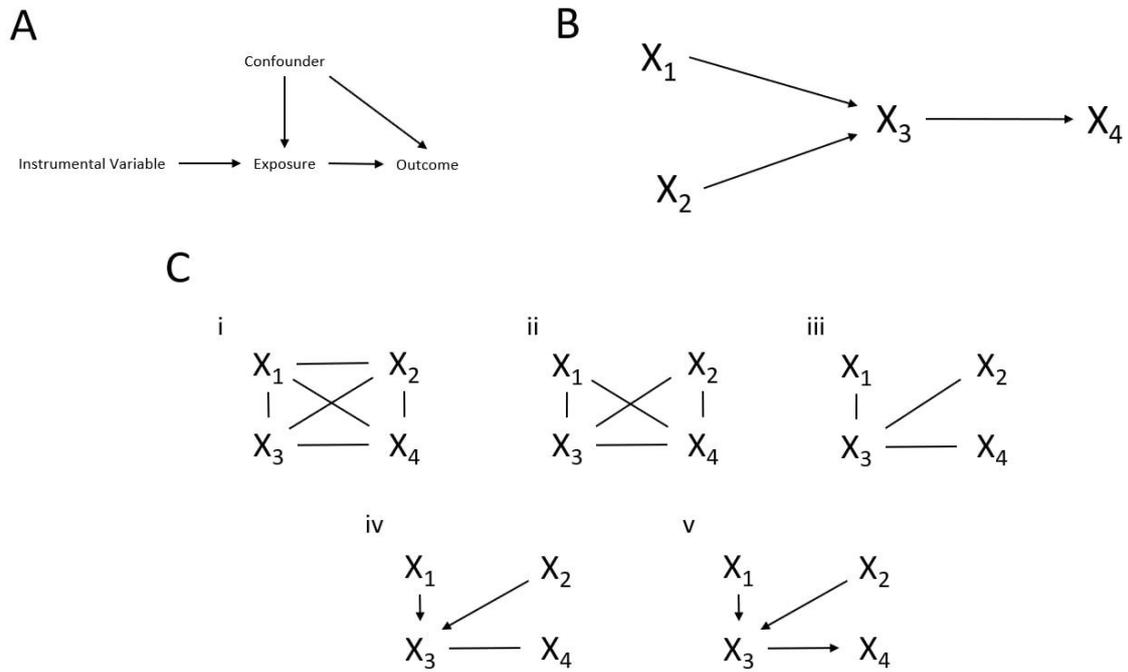

**Figure 2.** A) Schematic representation of MR. MR infers the causal effect of an exposure (phenotype) on the outcome using instrumental variables (IVs). B) Causal Bayesian networks connect nodes via directed edges determined by conditional independence, which is present when the relationship between two nodes is independent conditioning on all other nodes in the graph. C) Schematic representation of the PC algorithm. The true causal graph is shown in (B). The PC algorithm initially begins with an undirected fully connected graph (i) and uses data to create a skeleton graph with undirected edges. In this case, the $X_1$–$X_2$ edge is removed because $X_1$ is independent of $X_2$ (ii) and the edges between $X_1$–$X_4$ are removed as the nodes are independent given $X_3$. The same is true for the $X_2$–$X_4$ edge (iii). Then v-structures are identified (iv) and final edges oriented (v) (18).

*Table 1*. Summary of the discovery methods for analysis of causal molecular networks.

| Methodologies | Data source required | Advantages | Disadvantages |
| --- | --- | --- | --- |
| Mendelian randomisation (MR) | GWAS, omics | • Only requires summary statistics, fast to run<br>• Estimates causal effect size | • Data must meet certain (possibly untestable) assumptions<br>• Incapable of modelling complex relationships |
| Bayesian MR | GWAS, omics | • Flexibility of modelling complex data structure (overlapping samples, horizontal pleiotropy, interactions, multiple exposures)<br>• Estimates causal effect size | • Data must meet certain (possibly untestable) assumptions<br>• Computationally intensive, applicable to small-to-medium causal networks |
| Bayesian networks (BN) | omics | • Can generate larger causal networks<br>• Causal edges probabilities are given<br>• Estimates causal effect size | • Computationally intensive limiting network size |
| PC algorithm (18) | omics | • Relatively fast compared to other BNs | • Although faster than alternatives, computationally challenging when run on very large datasets<br>• Causal effect size is not inferred |
| Genome Granularity DAG (GDAG) (19) / findr (20)/ MRPC (21) | GWAS, omics | • Undirected network construction followed edge directions inferred using MR | • Still computationally intensive and applications have been on subsets of omics data<br>• Causal effect size is not inferred |
| Causal Graphical Analysis Using GEnetics (cGAUGE) (22) | GWAS, omics | • Approach has greater power and lower false discovery rate than BNs | • Computationally intensive<br>• Causal effect size is not inferred<br>• Greater power than BN, reduced power in presence of horizontal pleiotropy |
| Granger causality (23) | Time series omics | • Allows causal inference using time series omics data | • Time intervals between measurements needs to be enough for a noticeable change to take place<br>• Needs to be no confounders |
| Optimal Causation Entropy (OCE) (24)/ PCMCI (25) | Time series omics | • Outperform Granger causality using time series omics data<br>• Can generate large scale causal networks | • Assumes stationarity which can be violated by confounders |

*Mendelian randomisation*

MR uses single-nucleotide polymorphisms (SNPs) as 'instrumental variables' (IVs) to infer the causal effect of an exposure on an outcome. It mimics randomized controlled trials by assuming that SNP genotypes are randomly assigned to individuals within a population. MR has three key assumptions (Figure 2A) that must be met to infer causal relationships; a) IVs are associated with the exposure of interest; b) IVs are independent of confounders (both observed and unobserved) between exposure and outcome; c) IVs only affects the outcome through the exposure of interest.

Horizontal pleiotropy occurs when the IV influences outcome outside of its effect on the exposure, breaking the assumption that genotype only affects the outcome through the phenotype of interest. Several adaptations of MR have been developed to reduce the impact of horizontal pleiotropy. Popular approaches include MR-Egger (19) (which models pleiotropy assuming that effects of the IV on exposure and outcome are independent), MR-PRESSO (20) (which corrects for IVs with outlier effects) and Causal Analysis Using Summary Effect estimates (CAUSE) (21) (which accounts for correlated and uncorrelated pleiotropic effects). These newer approaches have seen wide adoption in the literature as they reduce these limitations to using MR. MR-PRESSO and MR-Egger are often both applied to data and results compared to reduce the impact of pleiotropy and outliers. These approaches have been used to provide evidence to support the casual effect of estimated glomerular filtration rate, a measure of kidney function, on chronic kidney disease, kidney stone formation, diastolic blood pressure and hypertension (22). Additionally, they have been used to show the causal effect of blood pressure on renal outcomes commonly affecting patients with hypertension (7).

In most cases, MR analysis requires the association between IV-exposure and IV-outcome are from two independent studies (23). This is known as two-sample MR. There are a limited number of one-sample MR methods that deal with IVs, exposures and outcomes coming from a single study (19,24). Some expansions to MR have been developed to handle data when two studies have overlapping individuals in common (25), which in classic MR approaches lead to bias. Zou *et al*. (26) have developed a more flexible Bayesian MR method that can handle one, two and overlapping samples. Bayesian MR has an advantage in its flexibility of coping with complex data structures, such as overlapping samples, horizontal pleiotropy, study heterogeneity and multiple exposure and outcomes, all in a single model (26–28).

More recent MR methods have included multiple approaches. Burgess *et al*. (29) presented their own contamination mixture method approach and Wang *et al*. (30) have proposed GRAPPLE (Genome-wide mR Analysis under Pervasive PLEiotropy), that utilises both strongly and weakly associated SNPs and can identify multiple pleiotropic pathways. Both have discussed the future importance of including multiple exposures in the study of genetics and MR.

Identifying IVs with similar causal estimates may improve identification of causal mechanisms. These advancements in MR methodologies provide researchers with more options to design models that better fit the assumptions of MR. Inferring causality using MR has been increasingly applied (31,32), however have been focused on smaller-to-medium scale and applications to large scale omics networks have been limited. Nevertheless, MR has found application being used in combination with other approaches to building molecular networks, which will be discussed shortly.

*Bayesian networks*

Bayesian networks (BNs) use Bayesian inference to calculate probabilistic graphical models of data. BNs are directed acyclic graphs (DAGs), a graph whose edges are directed and contain no bi-directional edges and there is no subset of nodes that can form a closed loop. The edges of the DAG are determined via conditional independence which is present when two nodes are independent conditioning on all other nodes in the graph. An example of a BN is shown in Figure 2B. There are two types of Bayesian networks, constraint-based and score-based. Constraint-based methods learn an undirected network skeleton using conditional independence testing and then assign the direction of edges between nodes that are not found to be independent. Score-based methods instead aim to optimise a scoring criterion across a search space of DAGs.

Bayesian networks were one of the first approaches proposed to investigate gene expression networks (5). Due to the high computational cost, most studies have been limited to inferring causal relationships within triplets of a gene regulatory network (33) with limited approaches to scaling networks to larger more complete molecular networks. Much of the literature using BN to infer molecular networks has introduced limitations to the size of the networks built. Mäkinen *et al*. (34) used BNs to investigate coronary artery disease, introducing genetic information as priors by not allowing genes that have no associated SNPs to be parents of genes that have an associated SNP. However, this was only done on a subset of genes rather than a full network.

Azad and Alyami (35) used BNs to investigate causal gene expression networks in Lapatinib resistance to better understand why some breast cancer patients have unsuccessful treatment. They used different Markov Chain Monte Carlo (MCMC) sampling algorithms to identify the optimum molecular network from the BN search space. MCMC samples a probability distribution where the next sample is dependent on the current sample. The study was limited to genes within the TGF-β signalling pathway in lapatinib sensitive and resistant breast cancer cells. They identified the driver genes as being associated with the GO biological terms positive regulation of pathway-restricted SMAD protein phosphorylation and regulation of lymphocyte.

Similarly, Bhattacharya and Das (36) used BNs to investigate causal genes in drug pathways for cancer, with a limited set of genes identified using machine learning and known drug target genes. Using a small dataset, they identified gene to gene connections that play a role in imatinib resistance in chronic myeloid leukaemia, including a *ACADVL* to *PDIA5* connection present in non-responder populations and not in those that respond to imatinib. These two proteins have been previously shown to play important roles in cancer drug-resistance (37). BNs have been used in the past to identify any causal effects of microRNA (miRNA) on gene expression interactions (38). However, these networks are very limited, with causal edges only from miRNA to gene expression and in many cases failed to identify known gene-gene interactions from experiment-supported databases.

Identifying the optimal BN is very difficult, and many approaches have been proposed with the aim to improve this process within transcriptional networks (35). These improvements have only generally shown to be moderate and computationally intensive for generating large networks. Large amounts of information could be missed if only a subset of data is used to build causal networks which is generally the approach used with BN due to the high computational cost. It is possible to sacrifice accuracy of networks for speed using approximate solutions (39), however this is not guaranteed to make it possible to build networks using data that is as highly dimensional as omics data. To overcome this problem, BNs are being used in combination with approaches including MR to optimise construction of large-scale networks.

*PC algorithm*

The PC algorithm (40) (named after its initial authors, Peter Spirtes and Clark Glymour) is used to estimate Bayesian networks, starting with a fully connected undirected graph and recursively deleting edges based on conditional independence properties. This generates a completed partially DAG (CPDAG) which consists of both directed and undirected edges. The steps the PC algorithm takes to build causal networks are shown in Figure 2C. The PC algorithm is fast for high dimensional and sparse problems, which makes it more suited towards uses with molecular network data (41).

Zhang *et al.* (42) used the PC algorithm with gene expression data to identify conditional independence between pairs of genes to build gene regulatory networks. Le *et al.* (43) predicted the causal mRNA targets of miRNAs using a method named Intervention-calculus when the DAG is Absent (IDA) (41). IDA has been shown to have use in investigating the impact of regulators on gene expression (44) but has seen little practical use to investigate disease. Zhang *et al.* (45) applied the method to epithelial-mesenchymal transition and multi-class cancer datasets and results were validated by transfections experiments.

Zhang *et al*. (45) used the IDA approach to infer miRNA-mRNA pair interactions, and identified differences in causal effects between different conditions. They have used IDA to infer causality of long non-coding RNA (lncRNA) on mRNA within modules identified using WGNCA to identify lncRNAs in specific biological functions (46), an approach that has also since been used to investigate pan-cancer (44).

Despite being faster than alternatives, the PC algorithm is still slow when applied to high dimensional datasets, and so as data is integrated runtime will increase (18). The PC algorithm has seen limited use on its own in applications to molecular networks. However, it has been used more recently in combination with other approaches to infer causality in biological data.

*Ensemble approaches*

Research is trending towards the use of ensemble approaches to building causal molecular networks, with the aim to reduce the limitations of individual approaches and build more robust networks. MR, in particular, has been combined with other methods to help topologies and speed up construction of causal network by putting constraints on edge directions.

Yazdani *et al*. (47) proposed an approach to identifying causal networks named genome granularity DAG (GDAG). Initially, strong IVs are generated from phenotype SNP data across each chromosome independently. The structure of the undirected network for omics data is identified, and the principle of MR is used to determine the directionality of edges using the strong IVs generated previously. They have used this approach to investigate the network of metabolites (48,49).

Augmenting Bayesian networks with the principles of MR has become popular for building molecular networks (8). Wang *et al*. (50) have tried to address the computational limitations of BNs on large-scale transcriptome-wide networks using a tool they have named findr. They used the SNPs that are directly associated with gene expression, known as expression quantitative trait loci (eQTLs). For each gene, the most strongly associated eQTL is selected as the IV in inferring the pairwise causal relationships between all genes in the network. These edges are ranked and assembled into a DAG (51). This method is much more efficient and outperforms traditional ways of building BNs, though has rarely been practically applied in the literature.

Badsha and Fu (52) have developed MRPC, which incorporates the principle of MR into the PC algorithm. The principle of MR is generalised to account for a variety of causal relationships between SNPs and molecular phenotypes. MRPC begins by learning the graph skeleton using the PC algorithm with an online false discovery rate correction and any edges are oriented to point from SNPs to molecular phenotypes. MRPC then looks for v-structures in the network between any 3 nodes and uses the principle of MR to help orient edges. Although MRPC has been shown to be very effective for building molecular networks, there is still room to develop further. Within small to medium networks MRPC performs exceptionally, however for very high dimensional data as is common with multi-omics data, it is still computationally expensive and could be further optimised.

A recent paper by Zuber *et al*. (53) proposed a multivariable MR and Bayesian model averaging (MR-BMA) approach that can include information from many IVs using only summary statistics from genetic association studies. It assumes the proportion of true causal risk factors is sparse when compared with all risk factors, which they demonstrate is usually the case with metabolomics data. Using MR-BMA, they identify high density lipoprotein (HDL) cholesterol as a potential causal risk factor for age-related macular degeneration, supported by previous literature (54). This approach has also been used to identify Apolipoprotein B as key lipid risk factor for coronary artery disease (55). All the above methods using the principle of MR require that the three assumptions of MR are satisfied. As multi-omics data is large and complex, using MR to sidestep the problems of confounding and reverse causation is important for causal network inference.

Causal Graphical Analysis Using GEnetics (cGAUGE) has also been proposed to construct causal networks by Amar *et al*. (56). cGUAGE first identifies conditional independencies in the data that are used to identify IVs for downstream MR, and for the construction of large-scale networks, which is called ExSep. Initially, the skeleton is found using the PC algorithm. Edges between nodes are then oriented. If SNPs are marginally associated with a node $X_2$, but are independent of $X_2$ given another node $X_1$, then this is used as evidence that $X_1$ is causal of $X_2$. cGUAGE does not infer causal effect size, so there is a lot of future potential in integrating ExSep with MR and other approaches to infer the skeleton.

*Time series data*

Time series data provides the opportunity to investigate molecular networks across a biological process. Generating causal networks is made much more difficult with the problems that inherently come with this data type. Particularly, the time between measurements may be inconsistent or not reflect the rate of change that is being investigated, causal relations can greatly change over time and unmeasured confounding variables may be introduced. As multi-omics data becomes easier to generate, there has been an increased interest in using time-series data to investigate molecular networks (57).

The most common approach to identifying causality in time series molecular data is Granger causality which assumes that variable X Granger-causes Y if values of X provide information that is significant about the future values of Y (58). Heerah *et al*. (59) have proposed Granger-causal analysis of gene expression data that can handle irregularly-spaced bivariate signals. However, it has some limitations that become obvious when using multi-omics data. The time intervals between measurements needs to be enough for a noticeable change to take place and there needs to be no confounders. Both assumptions are rarely met with biological data. Stehr *et al.* (60) have used Siamese neural networks for causal inference in time series data, which gives the approximate probabilities between nodes. However, this approach has only been performed on balanced synthetic data and has yet to be shown to be effective in real unbalanced data.

Causal analysis of time series molecular data is still very limited. Although new methodologies are being developed in other research domains (61), there has been limited applications to molecular networks. Modern algorithms such as Optimal Causation Entropy (OCE) (62) and the PC algorithm with a conditional mutual information (MCI) test to reduce autocorrelation and control false positive rates (PCMCI) (63) have been shown to outperform Granger causality and be able to handle large scale networks. Applying these approaches to molecular networks would be an important step in progressing the analysis of time series causal molecular networks.

**Biological interpretation of networks**

Networks of connected genes can quickly become very complex, which severely limits biological interpretation, even in simple co-expression network (64). Nevertheless, even when interpreting simple networks it is important to distinguish between association and causality. Inappropriate use of causal language has been a particular problem in biological sciences in the past (65).

Causal molecular networks are often high dimensional. Many studies (35,36) have identified smaller subsets of genes they are interested in through previous knowledge of pathways or clustering of undirected networks before inferring causality, however this can miss out factors that may be relevant within the causal network but are not within the cluster or not identified by traditional univariate analysis. Alternatively, constructing a causal network and then clustering the nodes would identify any functionally close sets of variables that are likely involved in similar biological processes. Few published papers have carried out clustering within causal molecular network. As the size of these networks grow, clustering will become increasingly important to identify biological processes and important causal molecules within them.

An advantage of large causal molecular networks is drug discovery and repurposing. Previous approaches to identifying drugs have been focussed on correlating transcription signatures between disease and known drugs (66) however this approach generates drugs and therapeutic targets that rarely are further researched, and have not had much success in brining any new treatments to the clinic. Causal pathways allow for more in-depth identification of drug targets. Škrlj *et al*. (67) have developed Causal Network of Diseases (CaNDis) which uses causal protein-protein interactions to identify FDA-approved drugs that can impact particular diseases. A known drug pathway signature from databases such as CMap (68) can be matched to the causal network to impact a particular target. Causal networks can also be studied to identify upstream regulators of known disease targets that can be targeted using drugs. Unfortunately, these advancements have had little use in the literature and thus limited translation to the clinic. Further development of methodologies and additional work using these drug discovery tools when constructing molecular causal networks should be included in future research as they become more accessible.

Network visualisation is often one of the first steps once networks have been created. One of the advantages of network visualisation is the ability to better communicate the results to readers and colleagues without a full understanding of how results were generated. Appropriate visualisation therefore becomes crucial to reflect the results and get the most from the data. There are many tools that assist in generating networks, including Cytoscape (69) and Gephi (70). These tools generally include a large amount of customisability to visualise the network, particularly in automatically generating layouts.

However, visualising and interpreting very large and complex networks can be difficult and often overlooked in the literature. Selecting the best and most appropriate way of displaying networks is very dependent on the type of network that is being visualised, and so requires a large amount of input by someone who understands the data and how it has been analysed. In molecular networks with multi-omics data, layering the different omics types within the visualisation to show how they interact would give a much more structured view than any predesigned layout that is available. Some approaches, including Bayesian networks and MR, provide causal effect sizes which can be visualised within networks by increasing size of edges for larger effect sizes. This allows experts from other biological fields to interpret the interactions of molecular phenotypes and is more likely to lead to future research. There is potential for creating interactive networks where nodes and edges can be included or excluded by adjusting a causal effect size threshold. One of the aims of causal inference is the identification of a small number of targets for therapeutic interventions and so effective visualisation with easy interpretation can be used by other researchers to identify networks of their particular interest.

**Conclusion**

Building causal molecular networks is becoming increasingly important in systems biology. Inferring causality from entirely observational data is much less time consuming and less expensive than traditional randomised trials or intervention experiments. Additionally, the availability of genetic and multi-omics data is massively increasing making casual molecular network inference a very powerful approach.

Here, we have reviewed the available approaches to building causal molecular networks. Traditional small-scale MR approaches infer causality between an exposure and outcome. This makes MR a powerful tool when combined with other approaches to build large-scale networks but very limited when used on its own. Bayesian network methods, including the PC algorithm, are based on conditional independence properties and rarely scale to large multi-omics networks well. Additionally, many of the methods developed based on Bayesian networks output a Markov equivalence class that may mean there may mean ambiguity between directed and undirected relationships.

Ensemble approaches to inferring causal networks have attracted increasing attention as they bring together the advantages of individual approaches, e.g. augmenting Bayesian networks with the principle of MR, such as MRPC (52) and findr (50). This has allowed for scaling of networks to a much larger size, however computational cost is still very high. Still, these approaches have not been widely applied in the literature and there is still much to improve. Reducing the impact of unmeasured confounders and horizontal pleiotropy is important in any complex causal inference and is why MR plays an important role in these approaches. These issues are being addressed with modern MR methods such as MR-egger (19), CAUSE (21) and Bayesian MR, and integrating these approaches into ensemble methods should be a focus in the future.

Selecting IVs is also a challenge for large-scale casual networks. Linkage disequilibrium and pleiotropic effects can violate IV assumptions. Selecting strong IVs would potentially reduce data size, thus reducing computation time, and reduce bias. However, there is a trade-off as only including strong IVs that only explain a small proportion of variation in the exposures may reduce the precision of the estimates. Therefore, the future challenge is to effectively identify and select for valid IVs that satisfy assumptions and are optimal for large causal molecular networks, which may prove to be especially difficult as it is not known if strong IVs will exist for every phenotype.

Many causal molecular network methods have focussed on use of individual level data, which can be difficult to get hold of as it is usually not included on public databases for ethical reasons. Improving the available methods that can infer causality using widely available summary statistics should be a priority for researchers so more can be done with current data. Improved methods, optimal interpretation and visualisation will advance understanding of disease processes. It is scientifically important but computationally challenging to take advantage of the increasing availability of multi-omics data that are now available, and directly translate to applications in clinical treatment of disease. Given the complex biological structure of certain outcomes, the literature points to a need to develop more flexible and comprehensive approaches to building causal molecular networks.

**Acknowledgements**

**Author Contributions**

JK and HG undertook the literature search and co-wrote the first draft and approved the final version. CB, BK and MT critically appraised the manuscript and approved the final version.

**Funding**

This work was jointly supported by the British Heart Foundation and The Alan Turing Institute (which receives core funding under the EPSRC grant EP/N510129/1) as part of the Cardiovascular Data Science Awards (Round 2, SP/19/10/34813).

**Ethical Approval**

No ethical approval was needed.

**Competing Interests**

The authors declare that they have no conflicts of interest.